\documentclass[aps,prl,twocolumn,amsmath,amssymb,amsfonts,superscriptaddress,floatfix,nofootinbib]{revtex4-1}
\usepackage{epsfig} \usepackage{graphicx}
\usepackage[pdftex,bookmarks]{hyperref}
\usepackage[pdftex]{color}
\usepackage[table]{xcolor}  
\usepackage{float} 
\usepackage{ulem} 

\newcommand{\bgar}{\begin{eqnarray}}
\newcommand{\enar}{\end{eqnarray}} 
 
 \newcommand{\be}{\begin{equation}}
\newcommand{\ee}{\end{equation}}  
 \def\mincirc{\lower
  3pt\hbox{$\buildrel<\over{\hbox{$\mathchar"218$}}$}}
\newcommand{\eotvos}{E$\ddot{\rm o}$tv$\ddot{\rm o}$s}

\begin{document}

\title{\bf Systematic errors in high-precision gravity measurements\\ by light-pulse atom interferometry on the ground and in space\\}

\date{\today}

\author{Anna M. Nobili}
\affiliation{Dept. of Physics ``E. Fermi'', University of Pisa, Largo B. Pontecorvo 3, 56127 Pisa, Italy}
\affiliation{INFN (Istituto Nazionale di Fisica Nucleare), Sezione di Pisa, Largo B. Pontecorvo 3,  56127 Pisa, Italy}

\author{Alberto  Anselmi}
\affiliation{Thales Alenia Space Italia, Strada Antica di Collegno 253, 10146 Torino, Italy}

\author{Raffaello Pegna}
\affiliation{Dept. of Physics ``E. Fermi'', University of Pisa, Largo B. Pontecorvo 3, 56127 Pisa, Italy}

\begin{abstract}
We focus on the fact that light-pulse atom interferometers measure the atoms' acceleration  with only three data points per drop. As a result, the measured effect of the gravity gradient is  systematically larger than the true one, an error linear with the gradient and quadratic in time almost unnoticed so far. We show how this error affects the absolute measurement of the gravitational acceleration  $g$ as well as ground and space experiments with gradiometers based on atom interferometry such as those designed for space geodesy, the measurement of the universal constant of gravity and the detection of gravitational waves.  When atom interferometers test the universality of free fall and the weak equivalence principle  by dropping different isotopes of the same atom one laser interrogates both isotopes and the error reported here cancels out. With atom clouds of different species  and two lasers of different frequencies the phase shifts measured by the interferometer differ by a large amount even in absence of violation. Systematic errors, including common mode accelerations coupled  to the gravity gradient with the reported error,  lead to hard concurrent requirements --on the ground and in space-- on several dimensionless parameters all of which must be smaller than the sought-for violation signal.
\end{abstract}

\maketitle

Light-pulse atom interferometers (AIs) are based on quantum mechanics. As the atoms fall, the atomic wave packet is split, redirected, and finally recombined via three atom-light interactions at times $0$, $T$, and $2T$. The phase that the atoms acquire during the interferometer sequence is proportional to the gravitational acceleration that they are subjected to. 

Although one might think that the phase shift  depends on quantum mechanical quantities ``\dots \textit{this is merely an illusion since we can write the scale factor} [between the phase shift and the gravitational acceleration] \textit{in terms of the parameters we control experimentally, i.e. Raman pulse vector $k$ and pulse timing $T$. It then takes the form $kT^{2}$. \dots We can simply ignore the quantum nature of the atom and model it as a classical point particle that carries an internal clock and can measure the local phase of the light field}.''\,(\cite{PetersThesis1998}, Sec. 2.1.3). 
The same reference also demonstrates that,  in the case of a gravitational field with a linear gradient, both  the exact path integral approach and  the purely classical one lead to the same exact closed form for the phase shift and free fall acceleration measured by the AI, which is then expanded in power series of the local gravity gradient $\gamma$ for convenience\,\cite{PetersThesis1998}. 
The recoil velocity is the only part of the atom-light interaction which is not found in the classical model; however, it does not appear in the phase shift actually measured by AIs, because they are operated symmetrically so as to cancel it out\,\cite{KasevichKreversal2006,Kasevich2018}.
Thus, the classical approach gives excellent predictions of the phase shift measured by the interferometer, while including the quantum mechanical details related to the internal degrees of freedom is needed to account for smaller effects, such as the finite length of the light pulses.

We focus on the fact that AIs measure the atoms' position along the trajectory only three times per drop (in correspondence with the three light pulses), unlike laser interferometers in falling corner-cube gravimeters which make  hundreds to a thousand measurements per drop\,\cite{Ashby2018}.  Hence, although predicted exactly, the gravitational acceleration measured by AIs is the true one only in a uniform field.

  The predicted value of the measured acceleration first appeared in Ref.\,\cite{StoreyCohenTannoudji1994}, and it has been confirmed ever since\cite{PetersThesis1998,WolfTourrenc1999,PetersMetrologia2001,Borde2001,Borde2003}. However,  none of these works mentions that in the presence of gravity gradient  this value is the \textit{average} free fall acceleration  (at time $T$ of the  middle pulse) based on three position measurements. This is, of course, only an approximation to the \textit{true }acceleration, expressed mathematically by the  second time derivative of the position, or obtained experimentally with  a sufficiently large number of measurements per drop.  

Initially, the lack of precise measurements made the difference unimportant, and by the time the precision improved nobody went back to this issue.
However, its physical consequences do deserve to be carefully addressed.
Using the classical approach\,\cite{PetersThesis1998} we point them out when AIs are used to measure the absolute value of the gravitational acceleration $g$,  for gravity gradiometry and for testing the universality of free fall (UFF), both on the ground and in space.

Since  UFF tests are included,  we allow from the start  the possibility that the equivalence of inertial and gravitational mass may be violated for atoms of different species $A,B$ in the field of Earth (violation of the weak equivalence principle, WEP) hence violating UFF\,\cite{NobiliAJP2013}. We therefore write the masses as  $m^{g}_{A,B}=m^{i}_{A,B}(1+\eta_{A,B})$, $M^{g}_{\oplus}=M^{i}_{\oplus}(1+\eta_{\oplus})$, where superscripts $i$, $g$ refer to inertial or gravitational mass and the \eotvos\ parameters   $\eta_{A}$, $\eta_{B}$ , $\eta_{\oplus}$ may not be exactly zero (although  they must be smaller than $1$ by many orders of magnitude\,\cite{TorsionBalanceFocusIssue,MicroscopePRL171204}). The equation of motion for atoms $A$ or $B$ reads
\begin{equation}\label{Eq:zdoubledotA,B}
\ddot z_{A,B}=-\frac{GM^{i}_{\oplus}}{(R_{\oplus}+z_{A,B})^{2}}(1+\eta_{\oplus}+\eta_{A,B})
\end{equation}
where $R_{\oplus}$ is the Earth's radius and the $z$ axis points upward. UFF is tested by measuring the differential acceleration $\ddot z_{B}-\ddot z_{A}$, then $\eta_{\oplus}$ cancels out, and there is a violation if, with identical initial conditions and no noise, the ratio
\begin{equation}\label{Eq:eta}
\frac{\ddot z_{B}-\ddot z_{A}}{(\ddot z_{A}+\ddot z_{B})/2}=
\eta_{B}-\eta_{A}\equiv\eta
\end{equation}
differs from zero; thus,  what matters is the different composition of the atoms under test, which should be maximized\,\cite{Adelreview2009,GGphaseA22009,MikePRL2013}. Hence we  assume $M^{g}_{\oplus}=M^{i}_{\oplus}\equiv M_{\oplus}$. 

Using a perturbative approach for  a gravity field  with a linear gradient  $\gamma$ (see, e.g., Ref.\,\cite{Book1987}) the equation of motion reads
\begin{equation}\label{Eq:zdoubledotA,BorderGamma1}
\ddot z_{A,B}\simeq -g_{\circ}(1+\eta_{A,B})-\gamma\Big(\frac{1}{2}g_{\circ}t^{2}-v^{\circ}_{A,B}t-z^{\circ}_{A,B}\Big),
\end{equation}
where $g_{\circ}=GM_{\oplus}/R_{\oplus}^{2}\simeq9.8\,\rm ms^{-2}$, $\gamma=2g_{\circ}/R_{\oplus}\simeq3.1\times10^{-6}\,\rm s^{-2}$, and $z^{\circ}_{A,B}$ and $v^{\circ}_{A,B}$ are the initial position and velocity errors of the  atoms at release (the exact values are assumed to be zero).
The solution is
\begin{equation}\label{Eq:Solution}
\begin{split}
 z_{A,B}(t)\simeq z^{\circ}_{A,B}+v^{\circ}_{A,B}t-\frac{1}{2}g_{\circ}(1+\eta_{A,B})t^{2}\\
 -\gamma t^{2}\Big(\frac{1}{24}g_{\circ}t^{2}-\frac{1}{6}v^{\circ}_{A,B}t-\frac{1}{2}z^{\circ}_{A,B}\Big)\ \ .
 \end{split}
\end{equation}
We compute the  phase shift $\delta\phi_{A,B}$ measured by the AI following the step-by-step algorithm  outlined in Ref.\,\cite{PetersThesis1998}. Assuming  the same   $k$ for all three pulses ($\hbar k$ is the momentum transfer, with $\hslash$ the reduced Planck constant)  and the same time interval $T$ between subsequent pulses, it is
\begin{equation}\label{Eq:PhaseShift}
\begin{split}
\delta\phi_{A,B}=\phi_{A,B}(2T)-2\phi_{A,B}(T)+\phi_{A,B}(0)\\
=k[z_{A,B}(2T)-2z_{A,B}(T)+z_{A,B}(0)]
\end{split}
\end{equation}
and, using (\ref{Eq:Solution})
\begin{equation}\label{Eq:PhaseShift2}
\begin{split}
\delta\phi_{A,B}(T)\simeq -kT^{2} \Big[g_{\circ}(1+\eta_{A,B})\\
+\gamma\Big(\frac{7}{12}g_{\circ}T^{2}-v^{\circ}_{A,B}T-z^{\circ}_{A,B}\Big)\Big].
 \end{split}
\end{equation}
With the scale factor $kT^{2}$ ($k$ and $T$  measured experimentally), this gives the free fall acceleration $g_{A,B\,meas}(T)$ that the AI is predicted to measure  at  time $T$ of the middle pulse. In modulus,
\begin{equation}\label{Eq:AccMeasuredGround}
\begin{split}
g_{A,B\,meas}(T)\simeq g_{\circ}(1+\eta_{A,B})
+\gamma\Big(\frac{7}{12}g_{\circ}T^{2}-v^{\circ}_{A,B}T-z^{\circ}_{A,B}\Big).
 \end{split}
\end{equation}
If $\eta_{A,B}=0$ (WEP and UFF hold) this is the same as in\,\cite{PetersThesis1998}; it is  the expansion to order $\gamma$ of an exact result which can be obtained in closed form by an exact path integral treatment or within a purely classical description.  

In a gravitational field with a linear gradient the free fall acceleration of the atoms at time $T$ is obtained from (\ref{Eq:zdoubledotA,BorderGamma1}) while the AI measurement gives, at the same time $T$, the value (\ref{Eq:AccMeasuredGround}), which is systematically larger (in modulus) than the true one  by the amount:
\begin{equation}\label{Eq:SystemticAccelerationError}
\Delta a=\frac{1}{12}\gamma g_{\circ}T^{2}\ \ 
\end{equation}
with a relative error  $\frac{\Delta a}{g_{\circ}}=\frac{1}{12}\gamma T^{2}$.  The discrepancy was pointed out in Ref.\,\cite{NobiliPRA2016} where it was  explained with the simple algebra involved in computing   (\ref{Eq:PhaseShift2}) from (\ref{Eq:PhaseShift}) and (\ref{Eq:Solution}). 
In physical terms, it is due to the limitation, intrinsic to the AI  instrument, of making only three position measurements  per drop.
Whether it can be neglected or not will depend on the specific experiment. If not, appropriate systematic checks are required in order to partially model this term, while  any remaining unknown fraction of it  must be too small to matter.

With the same perturbative approach the calculation can be extended to order $\gamma^{2}$ by  using  the solution $z_{A,B}(t)$ to first order in $\gamma$ as given by (\ref{Eq:Solution}), rather than to order zero (i.e. $z_{A,B}|_{\gamma=0}=-\frac{1}{2}g_{\circ}t^{2}+v^{\circ}_{A,B}t+z^{\circ}_{A,B}$), as used  in (\ref{Eq:zdoubledotA,BorderGamma1}). The new equation of motion  reads
\begin{equation}\label{Eq:zdoubledotA,BorderGamma2}
\begin{split}
\ddot z_{A,B}
\simeq -\Big[g_{\circ}(1+\eta_{A,B})+\gamma\Big(\frac{1}{2}g_{\circ}t^{2}-v^{\circ}_{A,B}t-z^{\circ}_{A,B}\Big)\\
+\gamma^{2}\Big( \frac{1}{24}g_{\circ}t^{4}-\frac{1}{6}v^{\circ}_{A,B}t^{3}-\frac{1}{2}z^{\circ}_{A,B}t^{2}  \Big)\Big]\ .
\end{split}
\end{equation}
Its  solution  leads to the phase shift,  hence to the acceleration measured by the AI to order $\gamma^{2}$:  
\begin{equation}\label{Eq:AccMeasuredGroundGamma2}
\begin{split}
g_{A,B\,meas}(T)\simeq g_{\circ}(1+\eta_{A,B})\\
+\gamma\Big(\frac{7}{12}g_{\circ}T^{2}-v^{\circ}_{A,B}T-z^{\circ}_{A,B}\Big)\\
+\gamma^{2}\Big(\frac{31}{360}g_{\circ}T^{4}-\frac{1}{4}v^{\circ}_{A,B}T^{3}-\frac{7}{12}z^{\circ}_{A,B}T^{2} \Big).
 \end{split}
\end{equation}
The result is the same as Eq.\,(2.19) in Ref.\,\cite{PetersThesis1998} (where it was obtained by expanding to second order the exact result in closed form) and it is generally accepted. However, it differs from the true acceleration given by (\ref{Eq:zdoubledotA,BorderGamma2}) (at the same time $T$) with a relative systematic  error:
\begin{equation}\label{Eq:SystematicErrorGamma2}
\begin{split}
\frac{\Delta a_{A,B}}{g_{\circ}}=\frac{1}{12}\gamma T^{2}
+\gamma^{2}\Big(\frac{2}{45}T^{4} -\frac{1}{12}\frac{v^{\circ}_{A,B}T^{3}}{g_{\circ}}-\frac{1}{12}\frac{z^{\circ}_{A,B}T^{2}}{g_{\circ}} \Big),
 \end{split}
\end{equation}
though  we limit our analysis to order $\gamma$.

Let us now consider an AI experiment in space, inside a spacecraft in low Earth orbit such as the International Space Station (ISS). The ISS is Earth pointing, the AI axis  is aligned with the radial direction and the nominal point $O$ of atoms' release  (origin of the radial axis $\zeta$ pointing away from  Earth) is at distance $h$ from the center of mass of the spacecraft (e.g., closer to Earth than the center of mass itself). When testing UFF with atom species $A$ and $B$  we can assume $\eta_{spacecraft}=\eta_{\oplus}=0$ since they cancel out anyway\,\cite{EPwithlaserranging}.  The equation of motion reads
\begin{equation}\label{Eq:zdoubledotA,Bspace}
\begin{split}
m^{i}_{A,B}\ddot\zeta_{A,B}=-\frac{GM_{\oplus}m^{i}_{A,B}}{[r-h+\zeta_{A,B}]^{2}}(1+\eta_{A,B})\\
+m^{i}_{A,B}n^{2}(r-h+\zeta_{A,B})
\end{split}
\end{equation}
with $r$ being the orbital radius of the spacecraft (constant for simplicity) and $n$ being  its orbital angular velocity  obeying  Kepler's third law $n^{2}r^{3}=GM_{\oplus}$. Since $(h-\zeta_{A,B})/r\ll1$, we can write
\begin{equation}\label{Eq:zdoubledotA,Bspace2}
\ddot\zeta_{A,B}\simeq-(a_{tide}+g_{orb}\eta_{A,B})+\gamma_{orb}\zeta_{A,B}         
\end{equation}
where $a_{tide}=\gamma_{orb}h$  is the tidal acceleration at the nominal release  point and  $g_{orb}=GM_{\oplus}/r^{2}\simeq8.7\,\rm ms^{-2}$, $\gamma_{orb}=3g_{orb}/r\simeq3.8\times10^{-6}\,\rm s^{-2}$ are the gravitational acceleration and  gravity gradient of Earth (the numerical values refer to an orbiting altitude of $\simeq400\,\rm km$). 
This equation  shows that in orbit the largest acceleration  is the tidal one, with $\frac{a_{tide}}{g_{orb}}\simeq3\frac{h}{r}\ll1$ while the driving acceleration of  UFF violation is $g_{orb}$ (slightly weaker than in ground drop tests), meaning  that  when the  free fall accelerations of two atom species are subtracted a composition dependent  violation signal would be $g_{orb}\eta$, with $\eta=\eta_{B}-\eta_{A}$.
The violation signal, if any, is an anomalous acceleration in the same direction (and unknown sign) as the monopole gravitational attraction from the source body, in this case the radial direction to the Earth's center of mass. Hence, the equation of motion (\ref{Eq:zdoubledotA,Bspace2}) contains the tidal acceleration (due to gravity gradient) in the radial direction and not its transversal component\,\cite{BlaserTides2001,TidesGG2003}.
The ratio of the variable acceleration $\gamma_{orb}\zeta_{A,B}$  relative to the constant term  $a_{tide}$ is   $\frac{\gamma_{orb}\zeta_{A,B}}{a_{tide}}=\frac{\zeta_{A,B}}{h}\simeq\frac{1}{2}\gamma_{orb}T^{2}$, in analogy to the corresponding  ratio on the ground $\frac{\gamma z_{A,B}}{g_{\circ}}\simeq\frac{1}{2}\gamma T^{2}$. Note that $\gamma_{orb}$ is only slightly larger than $\gamma$ while it is expected that $T$ can be  several times larger in space than on the ground, because of near weightlessness conditions. This  is considered  the key motivation for moving the experiment to space, since it means, for a given free fall acceleration, a larger phase shift and hence higher sensitivity (as $T^{2}$). However, it also means a larger gradient effect (also as $T^{2}$). With this warning we proceed with a perturbative approach as on  ground. To order $\gamma_{orb}$, it is
\begin{equation}\label{Eq:zdoubledotA,BorderGammaSpace}
\begin{split}
\ddot\zeta_{A,B}\simeq -\Big[a_{tide} 
+g_{orb}\eta_{A,B}\\
+\gamma_{orb}\Big(\frac{1}{2}a_{tide}t^{2}-\Upsilon^{\circ}_{A,B}t-\zeta^{\circ}_{A,B}\Big)\Big]
\end{split}
\end{equation}
where $\zeta^{\circ}_{A,B}$ and $\Upsilon^{\circ}_{A,B}$ are  position and velocity errors at release and the last term is of order $\gamma_{orb}^{2}$ but cannot be neglected because the free fall acceleration  to be measured is of order $\gamma_{orb}$. We are led to the  measured acceleration (in modulus):
\begin{equation}\label{Eq:AccMeasuredSpace}
\begin{split}
a_{A,B\,meas}(T)\simeq a_{tide}+g_{orb}\eta_{A,B}\\
+\gamma_{orb}\Big(\frac{7}{12}\gamma_{orb}hT^{2}-\Upsilon^{\circ}_{A,B}T-\zeta^{\circ}_{A,B}\Big)
 \end{split}
\end{equation}
which, by comparison with its theoretical counterpart (\ref{Eq:zdoubledotA,BorderGammaSpace}) at the same time, shows a  systematic relative  error $\frac{\Delta a}{a_{tide}}=\frac{1}{12}\gamma_{orb}T^{2}$, similar to the ground experiment.

When testing UFF release errors result in position and velocity offsets between the two atom clouds which --because of gravity gradient-- give rise to a systematic differential acceleration error that mimics a violation signal\,\cite{NobiliPRA2016}. The effect of release errors is known to be a major issue in  all UFF experiments based on ``mass dropping'',  while it does not occur if the test masses oscillate around an equilibrium position, as in torsion balance tests or in the proposed Galileo Galilei (GG) experiment in space\,\cite{BraginskyPaper2017}.
As proposed by Roura\,\cite{Roura2017}, the effect of release errors coupled to the local gradient can be eliminated if the momentum transfer of the second laser pulse is modified by a small  quantity of order $\gamma$ such that the atoms fall as if they were moving in a uniform  field. On the ground the nominal value $k_{2}$ to be applied at the second pulse is $k_{2}=k+\Delta k_{2}= k+k\frac{1}{2}\gamma T^{2}$.
A residual acceleration  $\gamma_{res}(z^{\circ}_{A,B}+v^{\circ}_{A,B}T)$ remains if this value is not implemented  exactly (a successful reduction  $\gamma_{res}/\gamma\simeq10^{-2}$ has been reported\,\cite{Kasevich2018}):
\begin{equation}\label{Eq:PhaseShiftRouraOnGround}
\begin{split}
\delta\phi^{\Delta k_{2}}_{A,B}(T)\simeq -kT^{2}\Big[g_{\circ}(1+\eta_{A,B})\\
-\gamma_{res}(z^{\circ}_{A,B}+v^{\circ}_{A,B}T)
+\frac{1}{12}\gamma g_{\circ} T^{2})\Big].
\end{split}
\end{equation}
The acceleration term (\ref{Eq:SystemticAccelerationError}) remains too, in  which the gradient is  unaffected by whatever reduction has been achieved for  the previous one, as pointed  out in the Comment\,\cite{Dubetsky2018} and acknowledged by Roura\,\cite{Roura2018}.
This is inevitable because $\Delta k_{2}$  has been computed in order to nullify the effect of the local gradient on the  atoms whose motion is governed by (\ref{Eq:zdoubledotA,BorderGamma1}). Instead, the acceleration   measured by the AI and used for tuning the change $\Delta k_{2}$,  is affected by the  error (\ref{Eq:SystemticAccelerationError}) which cannot therefore be compensated. Indeed, attempts   to compensate it\,\cite{Dubetsky2019} are questionable  because compensation  would alter the free fall acceleration of the atoms and force it to equal a measured value which (already to first order in $\gamma$) is not fully correct.

The very fact that in proposing the gravity gradient compensation scheme Roura did not address the acceleration term (\ref{Eq:SystemticAccelerationError}) indicates  that the systematic error made by taking only three measurements per drop has not been recognized.

A similar approach in space leads to a residual gradient $\gamma_{orb-res}<\gamma_{orb}$ after applying\,\cite{Roura2017}, and to  the phase difference:
\begin{equation}\label{Eq:PhaseShiftRouraSpace}
\begin{split}
\delta\phi^{\Delta k_{2}}_{A,B\,orb}(T)\simeq -kT^{2}\Big[a_{tide}+g_{orb}\eta_{A,B}\\
-\gamma_{orb-res}(\zeta^{\circ}_{A,B}+\Upsilon^{\circ}_{A,B}T)
+\frac{1}{12}\gamma_{orb}a_{tide}T^{2}\Big]
\end{split}
\end{equation}
where  the  error given  by the last term contains $\gamma^{2}_{orb}$, but amounts to  $\frac{1}{12}\gamma_{orb}T^{2}$ relative to $a_{tide}=\gamma_{orb}h$, which is the quantity to be measured, and therefore cannot be ignored as hinted by Refs.\,\cite{Dubetsky2018,Roura2018}.

A previous approach to reducing gravity gradient and initial offset errors in a proposed  test of UFF  on the ISS was based on the idea of rotating the interferometer axis\,\cite{MuellerPRA2017}.
For a dedicated mission the idea of rotating the whole spacecraft has been proposed by Rasel's group as the key to reduce tidal effects\,\cite{RaselMoriond2019}. 
In both cases the authors invoke a similarity with MICROSCOPE space experiment\,\cite{MicroscopePRL171204}. 

In MICROSCOPE, the offset vectors between the centers of mass of the macroscopic test bodies --being due to construction and mounting errors-- are fixed with the apparatus and therefore follow its rotation at all time, allowing the main tidal effect to be distinguished from a violation signal during the offline data analysis of a sufficiently long run --this is not a mass dropping experiment (Ref.\,\cite{BraginskyPaper2017}, Sec.\,7). Instead, mass dropping tests with AIs require a huge number of drops  to reduce single shot noise, each one with its own initial conditions and mismatch vector between different atom clouds, and the assumption that all these vectors are fixed with the apparatus cannot be taken for granted.
The argument presented in Ref.\,\cite{MuellerPRA2017}  that the proposed instrument \textit{``has random but specified mismatch tolerances''} is a weak one.
 Being systematic, this error  must be below the target acceleration of the test in all drops; otherwise --should mismatch reversal not occur even in a small number of drops during the entire run (which is hard to rule out by direct measurement)-- the resulting average acceleration will be larger than the target, thus questioning the significance of a possible ``violation'' detection.  

 Roura's proposal\,\cite{Roura2017} is therefore  to be preferred, as long as the acceleration term (\ref{Eq:SystemticAccelerationError}) is recognized and dealt with, if necessary.
 
On the ground the  error (\ref{Eq:SystemticAccelerationError}) affects the absolute measurement of $g$. The best such measurement  has achieved  $\Delta g/g\simeq3\times10^{-9}$\,\cite{PetersNature1999,PetersMetrologia2001}, only about three times worse than obtained by the absolute gravimeter with free falling corner-cube and laser interferometry\,\cite{FG51995}. With $T=160\,\rm ms$, the acceleration $\frac{7}{12}\gamma g_{\circ}T^{2}$ in (\ref{Eq:AccMeasuredGround})
exceeds the target error and has  required  a series of \textit{ad hoc} measurements (drops from different heights) to be modelled  and reduced below the target.  Should it be possible to improve the sensitivity of the instrument by increasing $T$, and to reduce the gradient and its effect coupled to initial condition errors as proposed by Ref.\,\cite{Roura2017}, the error (\ref{Eq:SystemticAccelerationError}) would still remain and should be taken care of for  the absolute measurement of $g$ to be improved.

In gravity gradiometers, two  spatially separated AIs with atoms  of the same species  interrogated by the same  laser (hence $\Delta T$=$0$ and $\Delta k=0$) measure their individual free fall accelerations at their specific locations and compute their difference. The advantage is that the differential (tidal)  acceleration is less affected than $g$ by disturbances mostly in common mode, such as vibration noise. They  are used for geodesy applications, but also for the measurement of the universal constant of gravity $G$ and the detection of gravitational waves.
 On the ground, if the release points  $A$ and $B$ are separated vertically by $\Delta h$ ($A$ at the  reference level and $B$ higher  by $\Delta h$), the  differential acceleration is
\begin{equation}\label{Eq:GradiometerGroundTheory}
\begin{split}
|g_{B\,theory}-g_{A\,theory}|
\simeq\gamma\Delta h \\
+\gamma\Big(\frac{7}{8}\gamma\Delta hT^{2}+(v_{B}^{\circ}-v_{A}^{\circ})T+z_{B}^{\circ} -z_{A}^{\circ} \Big)
 \end{split}
\end{equation}
while the gradiometer measures:
\begin{equation}\label{Eq:GradiometerGroundMeas}
\begin{split}
|g_{B\,meas}-g_{A\,meas}|\simeq\gamma\Delta h \\
+\gamma\Big(\frac{49}{48}\gamma\Delta hT^{2}+(v_{B}^{\circ}-v_{A}^{\circ})T+z_{B}^{\circ} -z_{A}^{\circ} \Big)
 \end{split}
\end{equation}
with a systematic acceleration error proportional to $\gamma^{2}$ which cannot be neglected relative to the tidal acceleration measured by the gradiometer, the fractional error being $\frac{7}{48}\gamma T^{2}$.
In space, with the release point $A$ as in (\ref{Eq:zdoubledotA,Bspace}), and  $B$ at a radial  distance $\Delta h$ (farther away from Earth), the gradiometer would measure
\begin{equation}\label{Eq:GradiometerSpaceMeasured}
\begin{split}\delta\phi_{B}-\delta\phi_{A}\simeq kT^{2}\Big[ \gamma_{orb}\Delta h\\
+\gamma_{orb}\Big(\frac{7}{12}\gamma_{orb}\Delta hT^{2}+(\Upsilon_{B}^{\circ}-\Upsilon_{A}^{\circ})T +\zeta_{B}^{\circ}-\zeta_{A}^{\circ} \Big)\Big]
 \end{split}
\end{equation}
with a fractional systematic error $\frac{1}{12}\gamma_{orb}T^{2}$. The  error --like the physical quantity to be measured-- contains the gradient.  Therefore, depending on the target precision and accuracy of the experiment, \textit{ad hoc} independent  measurements are needed in order to model and reduce it below the target.

In tests of  UFF with AIs in which different atoms $A$ and $B$ are dropped ``simultaneously'', the individual phase shifts are measured and their difference $\delta\phi_{B}-\delta\phi_{A}$ is computed, to yield zero if no composition dependent effect is detected   (i.\,e., $\eta=\eta_{B}-\eta_{A}=\frac{{\ddot z_{B}-\ddot z_{A}}}{(\ddot z_{A}+\ddot z_{B})/2}=0$, UFF and WEP hold). 

Different isotopes of the same atom can be interrogated with the same laser. In this case $T$ is the same and the gradient term with $\frac{1}{12}$ or $\frac{7}{12}$  coefficient cancels out. 
Different atom species need different lasers and a requirement arises, for a given target $\eta$ of the UFF test, on the time difference $\Delta T$, as pointed out  by Refs.\,\cite{Dubetsky2018,Roura2018}.
%
However, the main problem with different lasers  is that different frequencies ($k_{A}\neq k_{B}$) result in widely different phase shifts measured by the interferometer   even in case of  perfect synchronization ($\Delta T=0$), zero gradient ($\gamma=0$), no noise and no violation.
For instance, using  $^{87}\rm Rb$ and $^{39}\rm K$, the fractional difference of the phase shifts is of the order of  
$\frac{k_{K}-k_{Rb}}{k_{K}}\simeq1.67\times10^{-2}$ ($k_{K}=4\pi/767\,\rm nm^{-1}$,  $k_{Rb}=4\pi/780\,\rm nm^{-1}$).
When seeking a violation signal many orders of magnitude smaller this is a major problem, which never occurred before in the long history of these experiments that goes back to Galileo. Taking into account only $k_{A}\neq k_{B}$ the difference of phase shifts reads
\begin{equation}\label{Eq:PhaseShiftsDifference}
\begin{split}
\delta\phi_{B}-\delta\phi_{A}\simeq-g_{\circ}T^{2}[({k}_{B}-{k}_{A})+({k}_{B}\eta_{B}-{k}_{A}\eta_{A})]\ ,
\end{split}
\end{equation}
or $-g_{\circ}T^{2}[({k}_{B}-{k}_{A})(1+\eta_{A})+{k}_{B}\eta]$. As mentioned earlier, there is a large term even if WEP holds, which makes this quantity hardly suitable to detect a tiny violation. 
Moreover, the \eotvos\ parameters $\eta_{A}$ or $\eta_{B}$ appear in addition to $\eta=\eta_{B}-\eta_{A}$ [see Eq.\,(\ref{Eq:eta})] mixed with the Raman vectors. 
This mixing disappears and violation is correctly expressed by $\eta$  if we use the ratio of phase shifts instead: 
\begin{equation}\label{Eq:PhaseShiftsRatioGroundNoErrors}
\frac{\delta\phi_{B}}{\delta\phi_{A}}\simeq\frac{{k}_{B}}{{k}_{A}}(1+\eta)\ \ .
\end{equation}

By  defining $k_{A}=\overline{k}_{A}+\Delta k_{A}$, $k_{B}=\overline{k}_{B}+\Delta k_{B}$ with $\overline{k}_{A}$, $\overline{k}_{B}$ being the exact values and $\Delta k_{A}$, $\Delta k_{B}$ being the respective experimental errors we get
\begin{equation}\label{Eq:PhaseShiftsRatioGround}
\begin{split}
\frac{\delta\phi_{B}}{\delta\phi_{A}}\simeq\frac{\overline{k}_{B}}{\overline{k}_{A}}\Bigg\{1+\eta
-\frac{\gamma_{res}}{g_{\circ}}\Big[(v^{\circ}_{B}-v^{\circ}_{A})T+(z^{\circ}_{B}-z^{\circ}_{A})\Big]\\
+2\frac{\Delta T}{T}-  \frac{\Delta k_{A}}{\overline{k}_{A}}+ \frac{\Delta k_{B}}{\overline{k}_{B}}+ \frac{\Delta a_{dm}}{g_{\circ}}   -2\times\frac{1}{12}\gamma T^{2}\frac{a_{cm}}{g_{\circ}}  \Bigg\},
\end{split}
\end{equation}
where we have included  gravity gradient and initial condition errors, the synchronization and Raman vector errors, and also perturbations resulting  in common mode accelerations $a_{cm}$  with inevitable differential residuals
 $\Delta a_{dm}$.  The most relevant and best studied is vibration noise\,\cite{Chen2014};  by using the same mirror  it is ideally common mode, but a differential residual remains due to imperfect rejection. Systematic errors must obey the conditions
\begin{equation}\label{Eq:RequirementsAllErrorsGround}
\begin{split}
\frac{\gamma_{res}}{g_{\circ}}\big[(v^{\circ}_{B}-v^{\circ}_{A})T+(z^{\circ}_{B}-z^{\circ}_{A})\big]<\eta,\\
\frac{\Delta a_{dm}}{g_{\circ}}<\eta,  \ \  \  \frac{a_{cm}}{g_{\circ}}<\frac{6}{\gamma T^{2}}\eta,\\
 \frac{\Delta T}{T}<\frac{\eta}{2},\ \  \ \frac{\Delta k_{A}}{\overline{k}_{A}}<\eta \  \ \  \frac{\Delta k_{B}}{\overline{k}_{B}}<\eta.
\end{split}
\end{equation}
The requirement on initial condition errors is very severe\,\cite{NobiliPRA2016} and must be relaxed by applying, for each species, an appropriate frequency shift at the second laser pulse\,\cite{Roura2017} in order to make the residual gradient $\gamma_{res}$ as small as possible. Gravity gradient is relevant because it couples also to common mode accelerations which would otherwise cancel out (also to higher order). Thus, vibration noise must meet the tight requirements (\ref{Eq:RequirementsAllErrorsGround}) in differential and in common mode. Note that after applying\,\cite{Roura2017}, the requirement on $a_{cm}$ is relaxed only by a factor $7$  because the residual (\ref{Eq:SystemticAccelerationError}) contains the actual gradient $\gamma$ and not the reduced one $\gamma_{res}$.

Concerning synchronization, the requirement (\ref{Eq:RequirementsAllErrorsGround}) comes from the fact that each phase shift grows as  $T^{2}$ times the leading free fall acceleration, hence the relative error $2\frac{\Delta T}{T}$ competes  with $\eta$. The issue has been faced by\,\cite{RaselEP2014}  in a  WEP test with $^{87}\rm Rb$ and $^{39}\rm K$, though only to a few parts in  $10^{7}$.
By chirping the lasers at a particular rate a wave acceleration was applied in order to compensate to some extent (for each species)  the leading acceleration of the atoms which gives a phase shift proportional to $T^{2}$.
Raman pulse errors must meet similar tight requirements, and it is envisaged to make use of frequency comb technology\,\cite{Chen2014}. 

In Ref.\,\cite{Chen2014} the authors propose a data analysis that would allow a violation term containing $\eta$ to be separated from vibration noise.
However, this is not the only violation term in their equations, due to  the mixing shown by (\ref{Eq:PhaseShiftsDifference}).
Violation appears only as $\eta$ in the ratio of the phase shifts, and (\ref{Eq:PhaseShiftsRatioGround}) shows beyond question that vibration noise, both in common and differential modes,  cannot be separated from $\eta$; $\frac{\Delta a_{dm}}{g_{\circ}}$  and $\frac{1}{6}\gamma T^{2}\frac{a_{cm}}{g_{\circ}}$ could be misinterpreted as a violation and therefore must be below the tight bounds (\ref{Eq:RequirementsAllErrorsGround}).

For a violation at level $\eta$ to be detected, it is necessary (i) that all errors are negligible with respect to $\eta$ and (ii) that the absolute value of the ratio of the Raman vectors  $\frac{\overline{k}_{B}}{\overline{k}_{A}}$  is measured to both precision and accuracy better than $\eta$, so as to distinguish  a deviation from the measured value at this tiny level. Instead, in  UFF tests with macroscopic test masses, the  physical quantity of interest is zero if $\eta=0$ (``null experiments''\,\cite{BraginskyPaper2017}). 

With $\eta$ already established at levels below $10^{-13}$, $10^{-14}$\,\cite{TorsionBalanceFocusIssue,MicroscopePRL171204}, the requirements (\ref{Eq:RequirementsAllErrorsGround})  are  challenging, and each one needs a specific challenging technology, all to be implemented together. Every  error  could be  a WEP violation and needs  specific systematic checks in order to be distinguished from it; all checks must have the  target sensitivity and therefore require a total integration time each\cite{NobiliPRA2016,Roura2017}.

In space the ratio of phase shifts reads
\begin{equation}\label{Eq:PhaseShiftsRatioInSpace}
\begin{split}
\frac{\delta\phi_{B-orb}}{\delta\phi_{A-orb}}\simeq\frac{\overline{k}_{B}}{\overline{k}_{A}}
\Bigg\{1+\frac{g_{orb}}{a_{tide}}\eta\\
-\frac{\gamma_{orb-res}}{a_{tide}}\Big[(\Upsilon^{\circ}_{B}-\Upsilon^{\circ}_{A})T+(\zeta^{\circ}_{B}-\zeta^{\circ}_{A})\Big]\\
+2\frac{\Delta T}{T}-  \frac{\Delta k_{A}}{\overline{k}_{A}}+ \frac{\Delta k_{B}}{\overline{k}_{B}}+\frac{\Delta a_{dm}}{a_{tide}}-\frac{1}{6}\gamma_{orb}T^{2}\frac{a_{cm}}{a_{tide}}    \Bigg\};
\end{split}
\end{equation}
hence the requirements on systematic errors are: $\frac{\Delta T}{T}<\frac{\eta}{2}\frac{g_{orb}}{a_{tide}}$, $\frac{\Delta k_{A}}{\overline{k}_{A}}<\eta\frac{g_{orb}}{a_{tide}}$, $\frac{\Delta k_{B}}{\overline{k}_{B}}<\eta\frac{g_{orb}}{a_{tide}}$, $\frac{\gamma_{orb-res}}{g_{orb}}\big[(\Upsilon^{\circ}_{B}-\Upsilon^{\circ}_{A})T+(\zeta^{\circ}_{B}-\zeta^{\circ}_{A})\big]<\eta$, 
$\frac{\Delta a_{dm}}{g_{\circ}}<\eta$, $\frac{a_{cm}}{g_{orb}}<\frac{6}{\gamma_{orb}T^{2}}\eta$.
By comparison with (\ref{Eq:PhaseShiftsRatioGround}) and (\ref{Eq:RequirementsAllErrorsGround}),  the requirements on synchronization and Raman vector errors are relaxed by the large factor $\frac{g_{orb}}{a_{tide}}$, as noticed by\,\cite{Chen2014}, because in orbit the driving violation signal is $g_{orb}$ while the leading free fall acceleration is $a_{tide}$ [see Eq.\,(\ref{Eq:zdoubledotA,BorderGammaSpace})]. 
However, this factor is gained only for systematic  errors  linear with $a_{tide}$, not for gradient and initial condition errors and for vibration noise, both in common and differential mode, in which case the requirements are as tight as on ground.  


 The intrinsic limitations and severe requirements of UFF tests performed by dropping atoms of different  species are the reasons why almost all tests, especially if aiming at high precision\,\cite{Kasevich2007},  drop two isotopes of the same atom,   $^{87}\rm Rb$ and $^{85}\rm Rb$. However,  with only two neutrons difference, chances  are low that these experiments may detect composition dependent effects which would lead to new physics\,\cite{TorsionBalanceFocusIssue,ANAAmicroscope}.
 
Light-pulse atom interferometers  have the advantage that atoms provide both the test mass and  the readout. However,  they have only three time-position measurements each drop to recover the acceleration, unlike falling corner-cube gravimeters which can rely on hundreds to a thousand  data points per drop. The resulting systematic error grows linearly with the gradient and quadratically with the time interval $T$ between laser pulses. This error must be addressed in attempts to improve the absolute measurement of $g$ and must be proved to be irrelevant --or taken care of--  in gravity gradiometers for the measurement of the absolute value of the universal constant of gravity $G$, for space geodesy  and for the detection of gravitational waves.

This work has been supported by (ESA) Contract No. 4000125653 through an ITI type B grant. Thanks are due to Neil Ashby, Giuseppe Catastini,  Chris Overstreet, Marco Pisani and Massimo Zucco for useful discussions.

\end{document}